\documentclass[conference]{IEEEtran}
\IEEEoverridecommandlockouts
\usepackage{cite}
\usepackage{amsmath,amssymb,amsfonts}
\usepackage{algorithmic}
\usepackage{graphicx}
\usepackage{textcomp}
\usepackage{bm}
\usepackage{url}
\usepackage{xcolor}
\usepackage{url}
\def\BibTeX{{\rm B\kern-.05em{\sc i\kern-.025em b}\kern-.08em
    T\kern-.1667em\lower.7ex\hbox{E}\kern-.125emX}}
\begin{document}

\title{CompressedVQA-HDR: Generalized Full-reference and No-reference Quality Assessment Models for Compressed High Dynamic Range Videos}

\author{Wei Sun\textsuperscript{1}, Linhan Cao\textsuperscript{2}, Kang Fu\textsuperscript{2}, Dandan Zhu\textsuperscript{1}$^{\dag}$\thanks{$^{\dag}$Corresponding authors.},\\ Jun Jia\textsuperscript{2}, Menghan Hu\textsuperscript{1}, Xiongkuo Min\textsuperscript{2}, and Guangtao Zhai\textsuperscript{2}$^{\dag}$\thanks{This work was supported by the National Natural Science Foundation of China under Grants 62301316, 62377011, and 62225112.}\\
\textsuperscript{1}East China Normal University, \textsuperscript{2}Shanghai Jiao Tong University}


\maketitle

\begin{abstract}
Video compression is a standard procedure applied to all videos to minimize storage and transmission demands while preserving visual quality as much as possible. Therefore, evaluating the visual quality of compressed videos is crucial for guiding the practical usage and further development of video compression algorithms. Although numerous compressed video quality assessment (VQA) methods have been proposed, they often lack the generalization capability needed to handle the increasing diversity of video types, particularly high dynamic range (HDR) content. In this paper, we introduce CompressedVQA-HDR, an effective VQA framework designed to address the challenges of HDR video quality assessment. Specifically, we adopt the Swin Transformer and SigLip 2 as the backbone networks for the proposed full-reference (FR) and no-reference (NR) VQA models, respectively. For the FR model, we compute deep structural and textural similarities between reference and distorted frames using intermediate-layer features extracted from the Swin Transformer as its quality-aware feature representation. For the NR model, we extract the global mean of the final-layer feature maps from SigLip 2 as its quality-aware representation. To mitigate the issue of limited HDR training data, we pre-train the FR model on a large-scale standard dynamic range (SDR) VQA dataset and fine-tune it on the HDRSDR-VQA dataset. For the NR model, we employ an iterative mixed-dataset training strategy across multiple compressed VQA datasets, followed by fine-tuning on the HDRSDR-VQA dataset. Experimental results show that our models achieve state-of-the-art performance compared to existing FR and NR VQA  models. Moreover, CompressedVQA-HDR-FR won first place in the FR track of the Generalizable HDR \& SDR Video Quality Measurement Grand Challenge at IEEE ICME 2025. The code is available at \url{https://github.com/sunwei925/CompressedVQA-HDR}.
\end{abstract}


\section{Introduction}
\label{sec:intro}

Video compression is a fundamental process designed to reduce the storage and transmission demands of videos while striving to preserve their visual quality. With the rapid proliferation of video content across diverse platforms---ranging from traditional broadcasting, represented by standard-definition (SD), high-definition (HD), and standard dynamic range (SDR) videos, to online streaming services featuring a wide variety of content, such as user-generated content (UGC), high dynamic range (HDR) videos, 4K videos, and high frame rate videos---the role of video compression has become even more critical. As a result, accurately assessing the visual quality of compressed videos is essential not only for monitoring compression performance but also for guiding the development of more effective compression algorithms. 

In general, video quality assessment (VQA)~\cite{min2024perceptual} can be categorized into two paradigms: full-reference (FR)~\cite{zhang2018unreasonable,ding2020image,li2018vmaf} and no-reference (NR)~\cite{saad2014blind, korhonen2019two, li2019quality, wu2022fast}. FR VQA methods assess the ``quality difference" between the reference video (typically assumed to have perfect visual quality) and the distorted video, using this difference to determine the visual quality of the distorted video. In contrast, NR VQA methods directly estimate the visual quality of the distorted video without relying on any information from a reference video. Over the past two decades, a variety of VQA methods have been proposed to address the challenge of assessing the quality of compressed videos. For example, popular FR VQA methods such as PSNR, SSIM~\cite{wang2004image}, and VMAF~\cite{li2018vmaf} have been extensively employed for codec algorithm development and parameter optimization. Meanwhile, NR VQA methods like SimpleVQA~\cite{sun2022deep} and FAST-VQA~\cite{wu2022fast} have also demonstrated competitive performance compared to FR methods across several VQA benchmarks.


While these approaches have made notable progress, a persistent challenge remains: most existing methods exhibit limited generalization capability when confronted with the growing diversity of video content. For example, HDR videos significantly increase the bit depth used to represent each pixel. Since most VQA methods are primarily developed for conventional videos with SDR pixel formats, they often perform suboptimally on HDR content, making accurate visual quality assessment particularly difficult. To address this issue, we introduce \textbf{CompressedVQA-HDR}, a more generalized model designed for assessing the quality of compressed HDR videos. Our work builds upon CompressedVQA~\cite{sun2021deep}, a data-driven framework for both FR and NR VQA that has demonstrated strong performance on compressed UGC videos. Similar to its predecessor, CompressedVQA-HDR leverages deep neural networks (DNNs) to extract quality-aware features. For the FR model, we adopt the Swin Transformer~\cite{liu2021swin} as the backbone and compute deep structural and textural similarities between intermediate-layer features of the reference and distorted videos as the quality representation. For the NR model, we employ SigLip 2~\cite{tschannen2025siglip} as the backbone and extract the global mean of the final-layer feature maps as the quality feature. In both models, a two-layer multilayer perceptron (MLP) is used to regress the extracted features into final quality scores.

Given the relatively limited scale of existing HDR VQA datasets, directly training VQA models on such datasets poses a significant risk of overfitting, potentially resulting in poor generalization to out-of-distribution data. To address this challenge, we adopt a two-stage training strategy. For the FR model, we first pre-train it on a large-scale compressed UGC VQA dataset~\cite{ugc2021challenge} to learn generalized quality-aware representations, followed by fine-tuning on the HDRSDR-VQA dataset~\cite{ugc2025challenge}. For the NR model, we employ an iterative mixed-dataset training (IMDT) strategy, in which the model is trained simultaneously on multiple VQA datasets to improve its generalization capability across diverse video types. The NR model is then fine-tuned on the HDRSDR-VQA dataset. Experimental results demonstrate that the proposed models achieve state-of-the-art performance compared to existing FR and NR approaches, validating the effectiveness of our framework.

\section{Related Work}
\subsection{FR VQA Models}
FR VQA models assess visual quality by measuring the similarity or distance between the reference and distorted videos. Traditional metrics such as MSE and PSNR compute pixel-wise errors, whereas perceptual metrics like SSIM~\cite{wang2004image} and its extensions, including MS-SSIM~\cite{wang2003multiscale} and SSIMplus~\cite{rehman2015display}, evaluate structural similarities, which better align with the human visual system's perception of structural changes in images. Although these metrics were originally developed for images, they have been widely adopted in compressed video quality assessment studies. To more effectively capture temporal distortions, MOVIE~\cite{seshadrinathan2009motion} incorporates motion-tuned spatio-temporal analysis by computing quality along motion trajectories, thereby accounting for both spatial and temporal artifacts. VMAF~\cite{li2018vmaf}, a feature-fusion-based model, combines FR image quality assessment (IQA) features, with motion-related features derived from temporal frame differences, and employs a Support Vector Regressor to predict the final video quality scores.

Recently, with the advancement of DNNs, data-driven VQA models have been extensively explored. 
DeepVQA~\cite{kim2018deep} learns spatio-temporal visual sensitivity using a CNN-based architecture and employs a convolutional attention-based temporal pooling module to predict human-like video quality scores. 
C3DVQA~\cite{xu2020c3dvqa} employs a hybrid 2D/3D CNN to jointly learn spatial and temporal features from distorted and residual video frames, enabling end-to-end full-reference quality prediction that accounts for spatiotemporal masking effects. CompressedVQA-FR~\cite{sun2021deep} extracts deep structural and textural similarities~\cite{ding2020image} from intermediate CNN layers between reference and distorted frames, facilitating effective spatio-structural fidelity assessment for compressed UGC videos.

\subsection{NR VQA Models}
NR VQA models directly estimate visual quality from distorted videos without relying on reference content. Early NR VQA methods typically utilize handcrafted features such as natural scene statistics (NSS), low-level features, etc. to quantify distortion severity. For example, Video BLIINDS~\cite{saad2014blind} predicts perceptual video quality by extracting spatio-temporal natural scene statistics from frame-difference DCT coefficients and characterizing motion coherence and global motion. VIIDEO~\cite{mittal2015completely} computes local spatiotemporal statistics by applying frame differencing, divisive normalization, and asymmetric generalized Gaussian modeling to assess video quality without requiring any reference or training data. 

In recent years, data-driven NR VQA models have become the dominant paradigm. VSFA~\cite{li2019quality} extracts content-aware features using a pre-trained CNN and models temporal memory effects through a GRU network~\cite{cho2014learning} combined with a subjectively-inspired temporal pooling layer. SimpleVQA~\cite{sun2022deep} jointly learns spatial quality-aware features in an end-to-end manner and incorporates pretrained motion features, followed by MLP-based regression and multi-scale quality fusion guided by human visual perception. FAST-VQA~\cite{wu2022fast} adopts a fragment-based sampling strategy and introduces a Fragment Attention Network to achieve efficient end-to-end VQA. Minimalistic VQA~\cite{sun2024analysis} decomposes the VQA pipeline into four basic modules---a video preprocessor, a spatial quality analyzer, an optional temporal quality analyzer, and a quality regressor---all in minimalistic forms, yet achieving competitive performance. 

\subsection{HDR Related VQA datasets}
HDR VQA has emerged as a prominent research topic in recent years, with extensive efforts devoted to constructing subjective VQA datasets for HDR and its tone-mapped content. The LIVE HDR dataset~\cite{shang2023study} is the first publicly available large-scale HDR10 VQA dataset, comprising 310 videos with diverse content and distortion types, along with over 20,000 subjective ratings collected under two ambient lighting conditions. 
AVT-VQDB-UHD-1-HDR~\cite{rao2024avt} is an open-source UHD-1 HDR video quality dataset containing 195 compressed videos encoded with H.265, AV1, and VVC at multiple resolutions and bitrates, along with subjective MOS and objective model scores for benchmarking HDR VQA.
To support the evaluation of inversely tone-mapped (ITM) HDR videos, Zhou et al.~\cite{zhou2023dataset} introduced the ITM-HDR-VQA dataset, the first public benchmark comprising 200 ITM-HDR videos generated from 20 HDR scenes using 10 representative ITM algorithms, with subjective quality scores obtained through extensive psychovisual experiments. 
The LIVE HDRvsSDR dataset~\cite{ebenezer2024hdr} presents the first large-scale subjective and objective study comparing HDR and SDR videos of the same content under varying compression and resolution settings across different display devices, containing 356 videos with over 23,000 human ratings.

\section{Proposed Models}

\subsection{Video Preprocessing}
HDR videos typically feature high frame rates and high spatial resolutions, making direct processing of individual frames computationally expensive. To alleviate this issue, effective video preprocessing is essential. In this paper, we adopt a video preprocessing strategy similar to that used in CompressedVQA~\cite{sun2021deep}, where one frame is sampled per second, and the spatial resolution is downsampled to a level suitable for current DNNs. Specifically, given a distorted video $\bm{x}^d= \{\bm x^d_{i}\}_{i=0}^{N-1}$ and its reference one $\bm{x}^r= \{\bm x^r_{i}\}_{i=0}^{N-1}$, where $\bm x^d_{i}, \bm x^r_{i} \in\mathbb{R}^{H\times W \times 3}$ represents the $i$-th frame in distorted and reference video respectively. $H$ and $W$ are the height and width of each frame, and $N$ is the total number of frames. We first temporally downsample the raw video $\bm x_r$ and $\bm x_d$ from $R$ to $1$ fps, resulting in $\bm{z}^d= \{\bm z^d_{i}\}_{i=0}^{K-1}$, $\bm{z}^r= \{\bm z^r_{i}\}_{i=0}^{K-1}$, where $R$ is the original frame rate, $K=\lfloor N/R \rfloor$, $\bm z^d_{i} = \bm{x}^d_{\lfloor R\times i \rfloor}$, and $\bm z^r_{i} = \bm{x}^r_{\lfloor R\times i \rfloor}$. Following temporal sampling, each frame is spatially downsampled to a fixed resolution of $H' \times W'$ using bicubic interpolation, where $H' < H$ and $W' < W$. This ensures that the frames are appropriately resized to fit the input requirements of the VQA models while preserving essential visual content.

\begin{figure*}[htbp]
    \centering
    \includegraphics[width=0.68\textwidth]{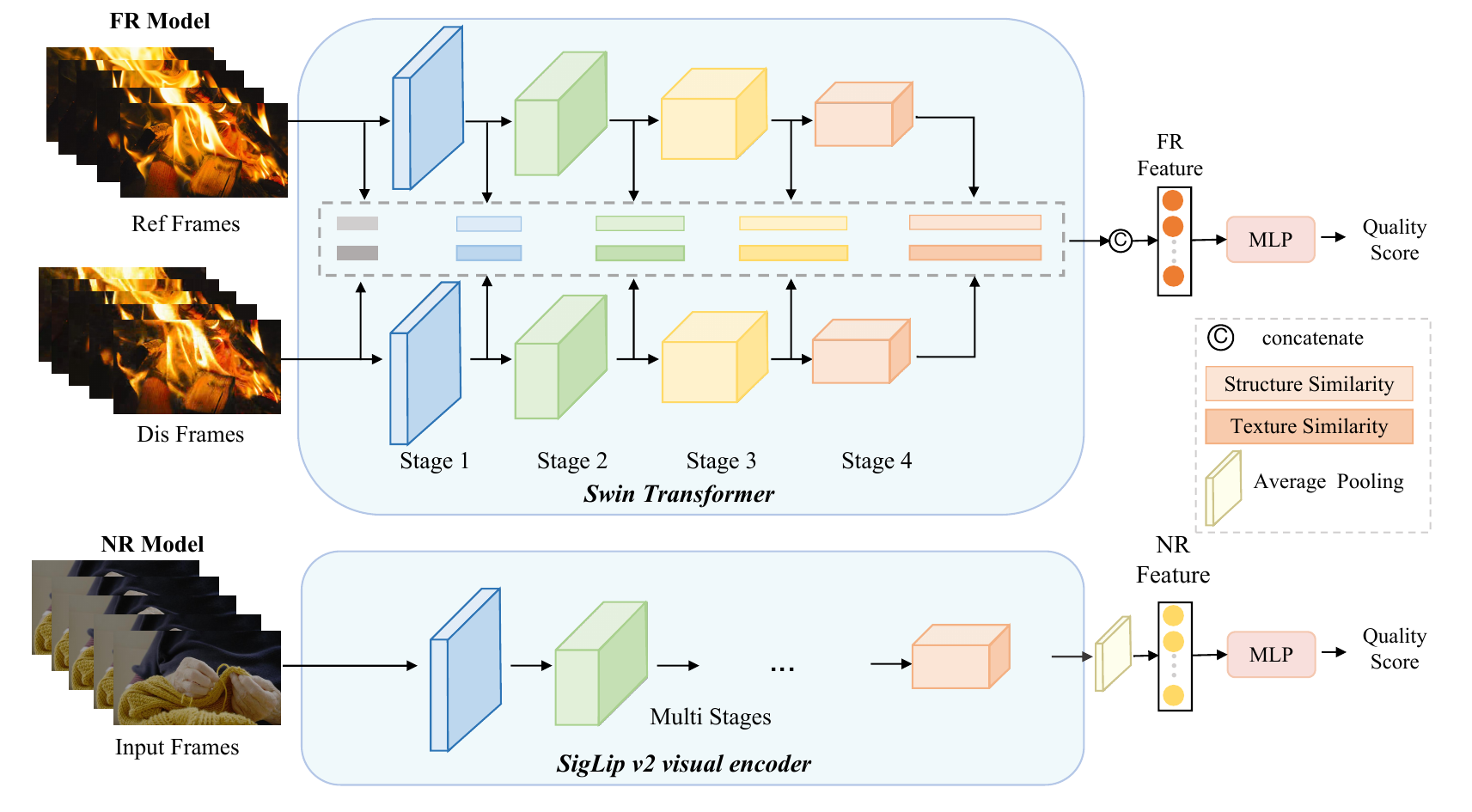}  
    \caption{Overall structure of our proposed FR and NR models.}
    \label{fig:model}
\end{figure*}

\subsection{Model Structures}
As illustrated in Figure~\ref{fig:model}, the proposed VQA models are composed of two main components: a quality-aware feature extraction module and a quality regression module.
\subsubsection{Quality-aware Feature Extraction Module}
For the FR model, we adopt a similarity computation strategy inspired by DISTS~\cite{ding2020image} and CompressedVQA~\cite{sun2021deep}, which measures the structural and textural similarity between deep features as a perceptual quality indicator. Specifically, we employ the Swin Transformer~\cite{liu2021swin} pretrained on the ImageNet dataset~\cite{deng2009imagenet} as the backbone network to extract deep features from video frames:
\begin{equation}
\begin{aligned}
\bm{f}^d_i = \mathrm{Swin}(\bm z^d_i), \quad \bm{f}^r_i = \mathrm{Swin}(\bm z^r_i),
\end{aligned}
\end{equation}
where $\mathrm{Swin}$ denotes the Swin Transformer model excluding the regression head. The resulting feature maps $\bm{f}^d_i$ and $\bm{f}^r_i$ are extracted from all stages of the Swin Transformer for the $i$-th distorted and reference frames, respectively. Then, the structure and texture similarities are defined as:
\begin{equation}
\begin{aligned}
\bm f_{T}\left(\bm{f}^r_i, \bm{f}^d_i\right) &=\frac{2 \bm \mu_{\bm{f}^r_i} \bm \mu_{\bm{f}^d_i}+c_{1}}{(\bm \mu_{\bm{f}^r_i})^2+(\bm \mu_{\bm{f}^d_i})^2+c_{1}}, \\
\bm f_{S}\left(\bm{f}^r_i, \bm{f}^d_i\right) &=\frac{2 \bm \sigma_{\bm{f}^r_i \bm{f}^d_i}+c_{2}}{(\bm \sigma_{\bm{f}^r_i})^2+(\bm \sigma_{\bm{f}^d_i})^2+c_{2}},
\end{aligned}
\end{equation}
where $\bm \mu_{\bm{f}^r_i}$ and $\bm \mu_{\bm{f}^d_i}$ denote the global means of the reference and distorted feature maps, $(\bm \sigma_{\bm{f}^r_i})^2$ and $(\bm \sigma_{\bm{f}^d_i})^2$ represent their global variances, and $\bm \sigma_{\bm{f}^r_i \bm{f}^d_i}$ denotes the global covariance between them. The constants $c_1$ and $c_2$ are small positive values added to ensure numerical stability. Finally, the structure and texture similarities computed at each stage of the Swin Transformer are concatenated to form the quality-aware feature representation of the FR VQA model, denoted as $\bm f_{\mathrm{FR}, i}$.

For the NR model, following the minimalistic VQA design principle in~\cite{sun2023analysis}, we directly use the output from the final layer of the DNN as the quality-aware feature representation. Specifically, we employ SigLip 2~\cite{tschannen2025siglip}, a high-performing vision-language model, as the backbone network. For each distorted frame $\bm{z}^d_i$, we compute the global mean of the feature maps from the last layer of the SigLip 2 visual encoder as follows:
\begin{equation}
\begin{aligned}
\bm{f}_{\mathrm{NR}, i} = \mathrm{GP_{mean}}(\mathrm{SigLip}(\bm z^d_i)),
\end{aligned}
\end{equation}
where $\mathrm{SigLip}$ denotes the visual encoder of the SigLip 2 model, $\mathrm{GP_{mean}}(\cdot)$ represents the global average pooling operation, and $\bm{f}_{\mathrm{NR}, i}$ is the resulting quality-aware feature representation for the NR VQA model.

\subsubsection{Quality Regression Module}
In this paper, we employ a two-layer multilayer perceptron (MLP) as the regression model to predict frame-level quality scores, consisting of 128 and 1 neurons in the first and second layers, respectively. The frame-level quality score is computed as follows:

\begin{equation}
\begin{array}{c}
\hat{q}_i = \mathrm{MLP}(\bm f_i), \bm f_i \in \{\bm f_{\mathrm{FR}, i}, \bm f_{\mathrm{NR}, i}\},
\end{array}
\end{equation}
where $\mathrm{MLP}$ denotes the two-layer MLP function, and $\hat{q}_i$ is the predicted quality score of the $i$-th distorted frame. The overall video quality score is then obtained by averaging the frame-level predictions:
\begin{equation}
\begin{aligned}
\hat{q}  &= \frac{1}{K} \sum_{t=1}^{K} \hat{q}_{i},
\end{aligned}
\end{equation}
where $\hat{q}$ represents the final video-level quality score predicted by the proposed model.

\subsection{Training Strategies}

Since existing HDR VQA datasets are relatively limited in scale, we propose training the model on multiple compressed VQA datasets to learn a more generalized quality representation. For the FR model, we adopt a transfer learning strategy: we first pretrain the model on a large-scale VQA dataset (i.e., the Compressed UGC VQA dataset~\cite{ugc2021challenge}), and then fine-tune it on the target dataset (i.e., the HDRSDR-VQA dataset~\cite{ugc2025challenge}).

For the NR model, we adopt an Iterative Mixed dataset Training (IMDT) strategy proposed by~\cite{sun2023blind}. This approach trains the model on multiple datasets simultaneously to enhance its generalization capability across diverse video content and distortion types. The quality-aware feature extraction module is shared across all VQA datasets, while individual quality regression modules are maintained for each dataset to accommodate differences in subjective quality scales. During training, the model iteratively optimizes each dataset-specific sub-problem, updating the shared feature extractor across all datasets and the corresponding regressor within each loop. 


Specifically, given $m$ VQA datasets $\{\mathcal{D}_i\}_{i=1}^m$, each containing distorted videos with subjective quality labels, we denote the shared quality-aware feature extraction module parameters as $\theta_s$, and the dataset-specific regression module parameters as $\theta_r^i$ for the $i$-th dataset. The objective for each dataset $\mathcal{D}_i = \{(x_j^i, q_j^i)\}_{j=1}^{N_i}$ is to minimize the loss function calculated by predicted and ground-truth quality scores:
\begin{equation}
\mathcal{L}_i(\theta_s, \theta_r^i) = \frac{1}{N_i} \sum_{j=1}^{N_i} \mathrm{Loss} ( f(x_j^i; \theta_s, \theta_r^i) - q_j^i ),
\end{equation}
where $f(\cdot; \theta_s, \theta_r^i)$ denotes the NR VQA model. During each training loop, we sequentially optimize the loss on each dataset by solving:
\begin{equation}
(\theta_s, \theta_r^i) \leftarrow \arg\min_{\theta_s, \theta_r^i} \mathcal{L}_i(\theta_s, \theta_r^i), \quad \text{for } i = 1, \dots, m,
\end{equation}
and update the shared parameters $\theta_s$ across all datasets while keeping each $\theta_r^i$ specific to its corresponding dataset.

To balance datasets with different sizes, the number of epochs $E_i$ allocated to each dataset within a loop is defined as:
\begin{equation}
E_i = \max\left( \left\lfloor \frac{N_{\max}}{N_i} \right\rfloor, E_{\text{min}} \right),
\end{equation}
where $N_i$ is the number of videos in $i-$th VQA dataset, $N_{\max} = \max( N_i )$, and $E_{\text{min}}$ is a predefined lower bound for training stability.


For the loss function, we use the PLCC loss to optimize the proposed FR and NR VQA models.

\section{Experiments}
\subsection{Training Datasets}
We use a combination of five representative VQA datasets for model training. The HDRSDR-VQA dataset~\cite{Chen2025HDRSDRVQA}, released by the ICME 2025 Grand Challenge, contains 360 videos (180 HDR and 180 SDR) generated from 20 open-source HDR source videos and their corresponding processed video sequences. The Compressed UGC VQA dataset~\cite{ugc2021challenge}, provided by the ICME 2021 Grand Challenge, includes 6,400 video clips for training and 800 for validation. Each reference video is compressed into seven distorted versions using H.264/AVC at various CRF levels. The WaterlooIVC-4K dataset~\cite{li2019avc} contains 1,200 compressed video sequences generated from 20 pristine 10-second 4K source videos using five modern encoders (AVC, HEVC, VP9, AVS2, and AV1), across three spatial resolutions and four distortion levels. The ITM-HDR-VQA dataset~\cite{zhou2023dataset} comprises 200 ITM-HDR videos derived from 20 HDR scenes using 10 representative inverse tone mapping algorithms. The LIVE Livestream dataset~\cite{shang2021assessment} includes 315 videos generated from 45 high-motion sports contents, each impaired by one of six common streaming-related distortions: compression, aliasing, flicker, judder, frame drops, and interlacing.

The FR model is first pretrained on the Compressed UGC VQA dataset and then fine-tuned on the HDR VQA dataset. For the NR model, we adopt an iterative mixed dataset training strategy using all the aforementioned datasets to learn generalized quality-aware features, followed by fine-tuning on the HDR VQA dataset. For the HDR VQA dataset, we divide the data into training and validation sets with an 8:2 split based on the reference videos. All experimental results are reported on the validation set.

\subsection{Evaluation Criteria}
We evaluate the performance of VQA models using four widely accepted metrics: Spearman Rank Order Correlation Coefficient (SRCC), Kendall Rank Order Correlation Coefficient (KRCC), Pearson Linear Correlation Coefficient (PLCC), and Root Mean Squared Error (RMSE). SRCC and KRCC assess the monotonic relationship between predicted and subjective scores, PLCC captures the linear correlation, while RMSE quantifies the overall prediction error.



\begin{table}
	\centering
	\renewcommand{\arraystretch}{1.15}
	\caption{The performance of the proposed FR VQA model and the compared methods on the HDRSDR-VQA dataset}
	\label{FR_Compressed}
	\begin{tabular}{c|cccc}
\hline
		
		Methods & SRCC & PLCC & KRCC & RMSE   \\
		\hline
		
		SSIM \cite{wang2004image}    &0.6269 &  0.6113&  0.4722 &1.0002  \\
		
		MS-SSIM \cite{wang2004image}    &0.6849 & 0.6573 & 0.5115  & 0.9525 \\
		LPIPS \cite{zhang2018unreasonable} & 0.6900	&0.6706 &0.5075 &0.9376\\
		DISTS\cite{ding2020image} & 0.6729&0.6546 & 0.5046& 0.9554\\
		VMAF \cite{xu2020c3dvqa} &  0.7358	&0.7081& 0.5657&  0.8925 \\
        ComprssedVQA \cite{sun2021deep} & 0.7814 	&0.7754&0.6039 &  0.7980 \\
		Proposed    & \textbf{0.9197} 	&\textbf{0.9348 } 	&\textbf{0.7498} 	&\textbf{0.4489} \\
		
		\hline
		
	\end{tabular}

	
\end{table}
\begin{table}
	\centering
	\renewcommand{\arraystretch}{1.15}
	\caption{The performance of the proposed NR VQA model and the compared methods on the HDRSDR-VQA dataset}
	\label{NR_Compressed}
	\begin{tabular}{c|cccc}
\hline
		
		Methods & SRCC & PLCC & KRCC & RMSE   \\
		\hline
		
		   NIQE~\cite{mittal2012making} & 0.0519 & 0.0103 & 0.0338  & 4.6467 \\
              SimpleVQA~\cite{sun2022deep}  & 0.8020 & 0.9039 & 0.6726   & 0.4796  \\
              FAST-VQA~\cite{wu2022fast}      & 0.7522 & 0.7133 & 0.5657  & 0.9570 \\
              DOVER~\cite{wu2023exploring}          & 0.8111 & 0.7985 & 0.6247  & 0.8023 \\
              MinimalisticVQA~\cite{sun2023analysis}              & 0.8810 & 0.8597 &  0.7097 & 0.6457 \\
              CompressedVQA ~\cite{sun2021deep}                 &0.8018 & 0.8605 &  0.6043 &  0.6440\\

		Proposed    & \textbf{0.9241} 	&\textbf{0.9261 } 	&\textbf{0.7663} 	&\textbf{0.4768} \\
		
		\hline
		
	\end{tabular}

	
\end{table}

\begin{table}
	\centering
	\renewcommand{\arraystretch}{1.15}
	\caption{Performance of the proposed FR VQA model and the baseline methods on the test set of ICME 2025 Grand Challenge}
	\label{FR_Compressed_challenge}
	\begin{tabular}{c|cccc}
\hline
		
		Methods & SRCC & PLCC & KRCC & RMSE   \\
		\hline
		
		  PSNR-Y  & 0.674& 0.677 &  0.480 & 1.015 \\
           VMAF &0.905 & 0.901 & 0.725  & 0.597 \\
		Proposed    & \textbf{0.932} 	&\textbf{0.941 } 	&\textbf{0.778} 	&\textbf{0.482} \\
		
		\hline
		
	\end{tabular}

	
\end{table}

\begin{table}[t]
    \centering
    \renewcommand{\arraystretch}{1.25}
    \caption{Ablation Studies of the Proposed FR models}
    \label{tab:ablation_FR}
    \begin{tabular}{c c|cccc}
    \hline
    Backbone & Pre-training   & SRCC & PLCC & KRCC & RMSE  \\
    \hline
   ResNet-50  & $\times$ & 0.8864 & 0.8579 &0.6990  &0.6493 \\ 
   Swin-B  & $\times$ & 0.9011 & 0.9029&  0.7275& 0.5432\\ 
   Swin-B  & \checkmark & \textbf{0.9197} 	&\textbf{0.9348 } 	&\textbf{0.7498} 	&\textbf{0.4489} \\ 
    \hline
    \end{tabular}
\end{table}

\begin{table}[t]
    \centering
    \renewcommand{\arraystretch}{1.25}
    \caption{Ablation Studies of the Proposed NR models}
    \label{tab:ablation_NR}
    \begin{tabular}{c c|cccc}
    \hline
    Backbone & IMDT   & SRCC & PLCC & KRCC & RMSE  \\
    \hline
   ResNet-50  & $\times$ & 0.8473 & 0.8470& 0.6538 & 0.6719\\ 
   Swin-B  & $\times$ &0.8763  & 0.8735&  0.6908& 0.6153\\ 
   SigLip-B v2  & $\times$ & 0.8924 & 0.9110& 0.7207 & 0.5211\\ 
   SigLip-B v2  & \checkmark & \textbf{0.9241} 	&\textbf{0.9261 } 	&\textbf{0.7663} 	&\textbf{0.4768} \\ 
    \hline
    \end{tabular}
\end{table}

\subsection{Compared Methods}
The compared FR VQA methods include SSIM~\cite{wang2004image}, MS-SSIM~\cite{wang2003multiscale}, LPIPS~\cite{zhang2018unreasonable}, DISTS~\cite{ding2020image}, VMAF~\cite{li2018vmaf}, and CompressedVQA-FR~\cite{sun2021deep}, while the compared NR VQA methods include NIQE~\cite{mittal2012making}, SimpleVQA~\cite{sun2022deep}, FAST-VQA~\cite{wu2022fast}, MinimalisticVQA~\cite{sun2023analysis}, DOVER~\cite{wu2023exploring}, and CompressedVQA-NR~\cite{sun2021deep}. For FR VQA models, we directly evaluate them on the validation set of the HDR VQA dataset. For NR VQA models, we retrain them on the training set before evaluating on the validation set.
\subsection{Training Details.}
We train the proposed models using the Adam optimizer with an initial learning rate of $1 \times 10^{-4}$ for the FR model and $1 \times 10^{-5}$ for the NR model, and a batch size of 6 on a server equipped with NVIDIA A800 GPUs. For the FR model, the pre-training stage is conducted for 10 epochs, followed by fine-tuning for 30 epochs. For the iterative mixed-dataset training strategy applied to the NR model, the minimum epoch threshold $E_{\mathrm{min}}$ is set to 10, with the number of training loops set to 3; during fine-tuning, the model is also trained for 30 epochs. The downsampled height and width of video frames, $H^\prime$ and $W^\prime$, are both set to 384.

\subsection{Experimental Results}
The experimental results of the FR and NR VQA models are presented in Table~\ref{FR_Compressed} and Table~\ref{NR_Compressed}, respectively. For FR models, we observe that existing FR VQA models achieve only moderate performance on the HDRSDR-VQA dataset, indicating their limited ability to evaluate HDR videos effectively. Among them, video-based methods such as VMAF and CompressedVQA outperform image-based metrics like SSIM and LPIPS; however, all these compared approaches are outperformed by the proposed FR model, highlighting the superior effectiveness of our method. For NR models, We observe that existing NR VQA models demonstrate varying levels of performance on the HDRSDR-VQA dataset. Traditional opinion-unaware metrics like NIQE yield poor results across all correlation metrics. Among the DNN-based approaches, DOVER and MinimalisticVQA exhibit competitive performance. Notably, the proposed NR model surpasses all existing methods, highlighting its superior effectiveness for HDR video quality assessment.

\subsection{Results on ICME Challenge}
We evaluate our FR model in the ICME 2025 Grand Challenge on Generalizable HDR and SDR Video Quality Assessment~\cite{ugc2025challenge}. The performance of our model, along with baseline models, on the challenge test set is presented in Table~\ref{FR_Compressed_challenge}. Our model achieves a performance gain of $2.98\%$ over the strong baseline VMAF in terms of SRCC, further demonstrating its effectiveness.

\subsection{Ablation Studies}
We investigate the effectiveness of the backbone network and the impact of the pre-training and IMDT techniques in Table~\ref{tab:ablation_FR} for the FR task and Table~\ref{tab:ablation_NR} for the NR task. The results show that Swin-B outperforms ResNet-50 in both FR and NR tasks, while SigLip-B v2 further surpasses Swin-B in the NR task, indicating that stronger backbone networks are more capable of capturing quality-aware features. Moreover, pre-training on the Compressed UGC VQA dataset for the FR model and IMDT on multiple datasets for the NR model lead to significant performance improvements, demonstrating that these strategies effectively transfer useful knowledge from other VQA domains to the HDR VQA task.

\section{Conclusion}
In this paper, we propose CompressedVQA-HDR, a unified framework for evaluating the visual quality of compressed HDR videos under both FR and NR settings. For the FR model, we compute deep structural and texture features between reference and distorted videos using Swin-B, while for the NR model, we extract the global mean of the final-layer features from SigLip-B v2, as their quality-aware representation. To address the challenge of limited training data, we adopt transfer learning for the FR model and iterative mixed-dataset training for the NR model to enhance generalization. Experimental results demonstrate that the proposed models achieve state-of-the-art performance on the HDRSDR-VQA datasets.

\bibliographystyle{IEEEbib}
\bibliography{icme2025references}

\vspace{12pt}

\end{document}